\documentclass[aps,prl,superscriptaddress,twocolumn]{revtex4}

\usepackage{graphicx}
\usepackage{amsmath}
\usepackage{color}
\usepackage{graphicx} 
\usepackage{dcolumn}  
\usepackage{bm}       

    \def\Rb{{\bf R}}  
\def\Eb{{\bf E}}  \def\pb{{\bf p}}  \def\kb{{\bf k}}

\begin{document}

\title{Enhanced microwave transmission through quasicrystal hole arrays}

\author{N. Papasimakis\footnote[1]{Electronic Mail: N.Papasimakis@soton.ac.uk}}
\address{Optoelectronics Research Centre, University of Southampton,
Southampton SO17 1BJ, United Kingdom}

\author{V. A. Fedotov}
\address{Optoelectronics Research Centre, University of Southampton,
Southampton SO17 1BJ, United Kingdom}

\author{F. J. Garc\'{\i}a de Abajo}
\address{Instituto de \'Optica - CSIC, Serrano 121, 28006 Madrid, Spain}

\author{A. S. Schwanecke}
\address{Optoelectronics Research Centre, University of Southampton,
Southampton SO17 1BJ, United Kingdom}

\author{N. I. Zheludev}
\address{Optoelectronics Research Centre, University of Southampton,
Southampton SO17 1BJ, United Kingdom}

\date{\today}

\begin{abstract}
We report on the observation of enhanced microwave transmission
through quasi-periodic hole arrays in metal films. The fraction of
transmitted light reaches $50\%$ in a self-standing metal film and
approaches $90\%$ when the film is sandwiched between thin
dielectric slabs, while the holes occupy only $10\%$ of the sample
area. The maximum transmission is accompanied by zero phase change,
rendering the film almost 'invisible' over a wide frequency range.
The extraordinary transmission phenomenon is interpreted in terms of
resonances in the self-consistent interaction between holes, which
are represented by effective electric and magnetic dipoles.
\end{abstract}

\maketitle


Little hope of having large light transmission through small
subwavelength apertures was allowed by the pioneering work of Bethe,
who predicted a drop in the intensity transmitted through a single
hole of radius $r$ in a thin perfect-conductor screen as
$(r/\lambda)^4$ for large wavelength $\lambda\gg r$ \cite{Bethe}.
However, the situation changed drastically when periodic arrays of
apertures were considered rather than isolated holes. Although hole
arrays have been extensively studied as artificial dielectrics in
the past \cite{chen}, the interest in this phenomenon was recently
renewed by the work of Ebbesen et al \cite{ebb98}, who demonstrated
experimentally that the optical transmission through subwavelength
hole arrays on metal films exceeds by several orders of magnitude
the original predictions by Bethe. Initially, the enhanced
transmission phenomenon was attributed to the excitation of surface
plasmons on the metal film surfaces \cite{ghaemi98, moreno01,
barnes04}, at variance with a dynamical diffraction interpretation
in that does not invoke plasmons and predicts the same effect in
perfect conductors \cite{treacy99, lezec04}. In fact, the same
effect was observed at THz \cite{thz,thz2,thz3} and GHz
\cite{mw1,mw2,mw3,mw4} frequencies, followed by the subsequent
identification of surface-bound modes in corrugated perfect
conductors \cite{pendry04,abajo05}, which play a similar role as
plasmons in the optical domain. Although the periodic pattern of the
structure was considered to be necessary to excite these surface
modes, extraordinary transmission was observed also in
quasi-periodic hole arrays in the optical regime
\cite{qc1,qc2,qc2b,qc3} and the origin of the phenomenon was traced
back to the interaction of surface plasmons with Bragg peaks in the
reciprocal space of the array \cite{qc1,qc3}, whereas similar
results were predicted also for perfect conductors \cite{qc4}.
\begin{figure}[!hb]
\centering
\includegraphics{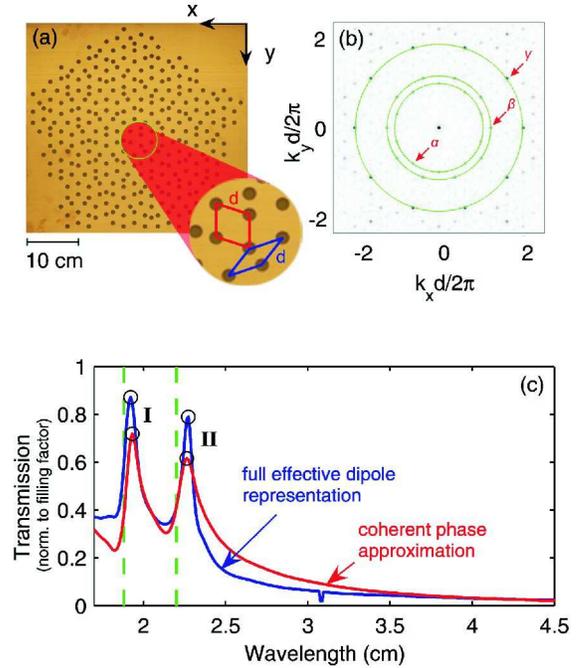}
\label{fig1} \caption{Figure 1. (a) Quasi-periodic hole array
drilled in a copper film deposited on a dielectric substrate. The
basic units of the quasicrystal are two rhombi with side length $d$
(inset). (b) Fourier transform of the quasi-periodic pattern
normalized to $d$. The three strongest Fourier maxima ($\alpha$,
$\beta$, and $\gamma$) are contained in the corresponding green
rings. (c) Normalized transmission of the hole array shown in (a),
calculated in the full dipole representation of Eq. \eqref{eq1}
(blue) and in the coherent phase approximation of Eqs. \eqref{eq2}
\& \eqref{eq3} (red). The dashed green lines mark the positions of
the Wood's anomalies.}
\end{figure}

It is known that a single hole on a thin \cite{Bethe} or a thick
\cite{abajo06} metal film can be represented by a magnetic dipole
parallel to the surface and an electric dipole perpendicular to it.
In the case of a hole array, the collective response admits a
representation in terms of the self-consistent polarization
$\pb_\Rb$ of each hole at the positions $\Rb$ in response to an
external field $\Eb^{\rm ext}$ plus the field induced by other holes
$\Rb'\neq\Rb$ via the hole polarizability $\alpha$, that is,
\begin{eqnarray}
   \pb_\Rb=\alpha [E^{\rm ext}(\Rb)+\sum_{\Rb'\neq\Rb}G(\Rb-\Rb') \pb_{\Rb'}],
   \label{eq1}
\end{eqnarray}
where $G(\Rb-\Rb')$ describes the field produced at hole $\Rb$ by
the polarization of the hole at $\Rb'$. In the small hole limit
$\lambda\gg r$, we can retain only the dipolar component of $\pb$
\cite{abajo06}. By considering an incident plane wave with
$\kb_\parallel$ momentum parallel to the film and assuming a
$\exp(i\kb_\parallel\cdot\Rb)$ spatial dependence for the hole
polarizability \cite{cpa}, Eq.\ \eqref{eq1} can be rewritten as
\begin{eqnarray}
   \pb_\Rb\approx\frac{1}{\frac{1}{\alpha}-G_{\kb_\parallel}}E^{\rm
   ext}(\Rb),
   \label{eq2}
\end{eqnarray}
where $G_{\kb_\parallel}=\sum_\Rb G(\Rb)
e^{-i\kb_\parallel\cdot\Rb}$ is the sum of the dipole-dipole
interaction over the quasi-lattice. Finally, the transmission $T$ is
given by the coherent superposition of the far field produced by all
induced dipoles, or equivalently, the transmission along a direction
defined by a projected parallel momentum $\kb_\parallel^{\rm out}$
is the far field produced by the dipole
\begin{eqnarray}
   \sum_\Rb \pb_\Rb e^{-i\kb_\parallel^{\rm
out}\cdot\Rb}.
   \label{eq3}
\end{eqnarray}
The lattice sum $G_{\kb_\parallel}$ exhibits pronounced maxima when
the main diffraction peaks become grazing, which are the equivalent
of the Wood anomaly condition in quasicrystal arrays. According to
Eq. \eqref{eq2}, the transmission will actually exhibit a minimum at
the divergences of $G_{\kb_\parallel}$ and a transmission maximum
signaled by the minimum value of $|1/\alpha-G_{\kb_\parallel}|$.

In order to further investigate the extraordinary transmission
mechanism, we consider a quasi-periodic pattern consisting of $313$
circular holes of radius $r=0.46~cm$. The side of the repeated basic
units of the array is $2.31~cm$ (see Fig. 1a). In Fig. 1b, the
reciprocal space of the quasicrystal is also shown, where it can be
seen that it is composed by dense Bragg peaks and exhibits high
orientational order. Although a very large number of peaks is
visible, three very strong Fourier maxima can be distinguished
($\alpha$, $\beta$, and $\gamma$), located at the circumference of
circles with dimensionless radii equal to $1.05$ ($\alpha$), $1.23$
($\beta$), and $1.98$ ($\gamma$), corresponding to spatial periods
of $2.20~cm$, $1.88~cm$ and $1.17~cm$, respectively. The relation
between the diffraction and transmission peaks becomes apparent in
the calculated transmission spectra presented in Fig. 1c, using the
coherent phase approximation (red) and the full solution of Eq.
\eqref{eq1} (blue). Both calculations coincide reasonably justifying
the coherent phase approximation. Moreover two transmission maxima
are predicted at $1.92~cm$ (I) and $2.27~cm$ (II), corresponding to
the two lowest frequency Fourier maxima of the quasicrystal
($\alpha$ and $\beta$).

The quasi-periodic pattern described above was used to manufacture
two different samples of $44~cm$ x $46~cm$ overall size, a self
standing aluminium film of $0.5~mm$ thickness and a $35~\mu m$
copper film residing on a $1.5~mm$ thick dielectric substrate with
permittivity $\epsilon=3.77+0.03i$ (see Fig. 1a). The microwave
measurements were performed in the range of $2~GHz$ to $18~GHz$, in
an anechoic chamber using a vector network analyzer and two horn
antennas. The sample was placed between the antennas and the
transmitted intensity and phase at normal incidence were recorded.

The results for the polarization along the y-axis (see Fig. 1a) are
shown in Fig. 2 and are normalized only to transmission through free
space. For the orthogonal polarization similar (although not
identical) results were obtained and are therefore omitted. In Fig.
2a, we present the data for the self-standing metal film. Two sharp
transmissions peaks can be seen at $2.02~cm$ (I) and $2.34~cm$ (II)
wavelengths on a slowly decaying transmission background. The
magnitude of the peaks is $50\%$ and $48\%$ respectively, while at
the same time the phase change of the transmitted wave for both
peaks is close to zero. When the metal film is supported by a
dielectric substrate (Fig. 2b), peaks I and II become considerably
weaker ($30\%$ and $31\%$) and are separated by a point of zero
transmission at $2.1~cm$, where the phase is undefined. Moreover, a
new transmission peak appears at $2.93~cm$ (III) (Fig. 2b). The
magnitude of the new peak is approximately $65\%$ and is accompanied
by a zero phase change. If, in addition to the dielectric substrate,
a superstrate of the same thickness and permittivity is introduced
(Fig. 2c), peaks I and II are no longer visible. On the other hand,
peak III increases in magnitude and reaches $90\%$, while the phase
at the maximum is again zero. Moreover, peak III becomes broader and
remains over $50\%$ over a wide frequency range, from $7.5~GHz$ to
$11~GHz$.
\begin{figure}[!h]
\label{f2} \centering
\includegraphics{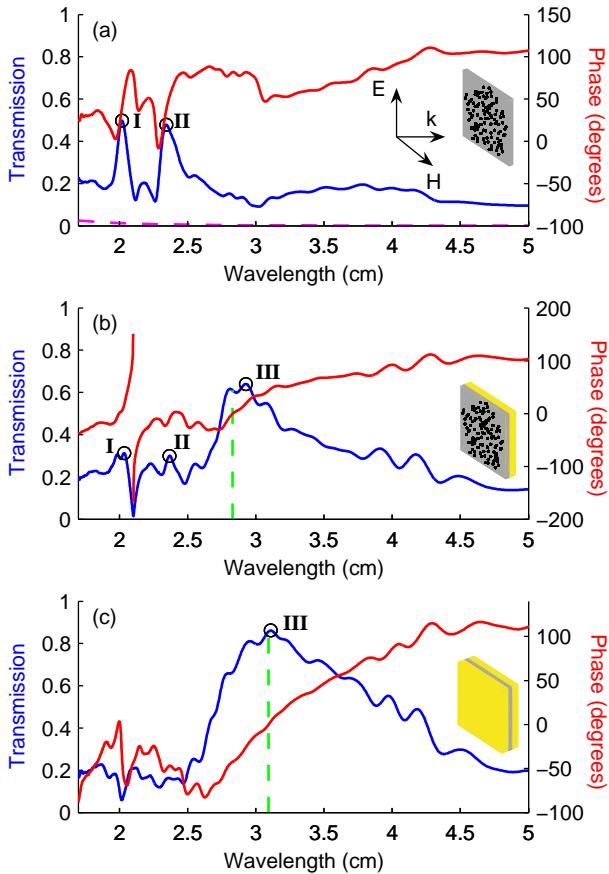}
\caption{Figure 2. Normal incidence transmission spectra through
quasicrystal hole arrays on (a) a self-standing Al film, (b) a
copper film supported by a dielectric substrate and (c) a copper
film sandwiched between two identical dielectric slabs. The Bethe
prediction is shown by the dashed purple line in (a), while the
green lines in (b) and (c) mark the wavelength positions of
"invisible metal" states, where high transmission is accompanied by
zero phase change.}
\end{figure}

In all cases, the transmission through the hole arrays exceeds
Bethe's predictions, since about $\sim10$ times more intensity than
what is directly incident in the area occupied by holes is
transmitted. The positions of the transmission peaks are dictated by
the Fourier maxima of the quasi-periodic pattern and the presence of
additional dielectric layers. In particular, for the case of the
self standing array (Fig. 2a), the structure is symmetric and
degenerate surface modes are excited along the the two metal-air
interfaces. The observed transmission peaks occur, as expected, very
close to the positions predicted by the theoretical curves of Fig.
1a, while the peaks corresponding to shorter spatial frequencies
($\gamma$) are not observed, since they lie out of the measured
frequency range. The more complicated spectral shape observed for
the non-symmetric structure of Fig. 2b can be explained by taking
into account the fact that the surface states on either side of the
metal film are no longer degenerate. The transmission peaks (I, II)
that originate from the metal-air interface are still visible, but
they become considerably weaker. On the other hand, the
dielectric-air interface leads to two new transmission peaks shifted
to longer wavelengths. However, this frequency shift results also in
an increase of the peak width and we believe that the two maxima
partially overlap forming a very broad peak at $\sim 3$ cm (III).
Furthermore, the confinement of the field near the metal surface
becomes stronger and maximum transmission increases to $65\%$, while
the corresponding phase change is zero, meaning that the incident
wave remains almost unaffected as it propagates through the
structure. Moreover, the phase singularity observed in between peaks
I \& II, could be attributed to the existence of Wood's anomalies in
the metal-dielectric interface, although further investigation is
required. When the degeneracy of the surface states is restored by
adding a superstrate, only the joint peak (III) survives, while at
the maximum the $90\%$ transmission and the zero phase change render
the structure virtually "invisible".

In conclusion, we have demonstrated, theoretically and
experimentally, enhanced transmission of microwaves through
quasi-periodic hole arrays in perfect conductors which can not
support surface plasmons and a direct relation between the
reciprocal space maxima and the transmission peaks was established.
In particular, an "invisible metal" state has been observed, where
almost total transmission with zero phase change can be achieved by
placing a structured film between two dielectric slabs. The
wavelength position of the total transmission can be tuned either by
varying the permittivity of the dielectric slabs or by appropriately
scaling the pattern. In fact, we have already shown that this design
is widely scalable and exhibits extraordinary transmission down to
the telecom spectral region \cite{qc2,qc2b}. Furthermore, the
results presented here are almost independent of the polarization of
the incident wave, due to the high orientational order of the
quasicrystal. These characteristics are much desired in practical
applications and we expect that such structures can prove useful
over a wide region of the electromagnetic spectrum.

\begin{acknowledgments}
The authors would like to acknowledge the financial support of the
Engineering and Physical Sciences Research Council, UK.
\end{acknowledgments}


\begin{thebibliography}{24}
\bibitem{cpa} The long-range interaction between holes dictates the
transmission resonances, so that local variations between
neighboring holes can be overlooked in this coherent-phase
approximation, where the phase imposed by the external field
reflects the effective phase difference in the response of distant
holes.
\end{thebibliography}
\end{document}